\documentclass[[11pt]{article}
\usepackage{russKOI8R}
\usepackage{iheprepeKOI8R}
\usepackage{ipsizen}

\usepackage{graphicx}
\usepackage{subfigure}

\def\m4g{m_{4 \gamma}}
\def\f1270{$f2(1270)$}

\begin{document}
\begin{titlepage}
\prepnum{2010-15} {ОЭФ}
\vspace{1.0cm}
\begin{center}

{\bf M.Yu.Bogolyubsky, Yu.V.Kharlov, D.I.Patalakha, B.V.Polishchuk, 
\\ S.A.Sadovsky, A.S.Soloviev, M.V.Stolpovskiy}
\vskip 0.2cm

\vskip 1.0cm
\end{center}

\title{\Large Correction of the energy scale nonlinearity \\ 
in electromagnetic calorimeters with the $\pi^0$ two-photon decays}
%

\end{titlepage}

\begin{abstractpage}[519.25.256]

\numref{4} 
 
\rusabs{М.Ю.Боголюбский и др. }{Коррекция нелинейности энергетической шкалы 
электромагнитного калориметра по двухфотонным распадам $\pi^0$-мезона.}

В работе представлен метод вычисления коррекции нелинейности отклика
электромагнитного калориметра, основанный на минимизации отклонения
измеренной массы нейтрального мезона, распадающегося в конечном счете 
на фотоны, в зависимости энергий последних. Метод был разработан и 
применён для электромагнитного калориметра LGD2 в эксперименте 
Гиперон-М на ускорителе У70 ГНЦ ИФВЭ. Найденная коррекция позволила
существенно уменьшить вариации реконструированных масс $\pi^0$ и $\eta$
мезонов в зависимости от их минимальной энергии.

\engabs{M.Yu.Bogolyubsky et al.}{Correction of the energy scale nonlinearity  
in electromagnetic calorimeters with the $\pi^0$ two-photon decays}

The method to calculate the non-linearity correction of the
electromagnetic calorimeter response, based on minimisation of the
deviation of the measured neutral meson mass on the energies of it
decay photons, is described in this paper. This method was developed 
for the electromagnetic calorimeter LGD2 in the Hyperon-M experiment  
at U70 accelerator of IHEP. The found correction allowed to reduce 
significantly variations of the reconstructed $\pi^0$ and $\eta$ 
masses on the minimal energy of the mesons.

\end{abstractpage}

\section*{Introduction} 

Photons and electrons due to interaction with a medium of the
cell-type electromagnetic calorimeter  
produce electromagnetic showers which spreads over several calorimeter
cells called a shower cluster, 
i.e. the group of affected cells with common edges. 
The read-out electronics for such kind of calorimeters reads 
the signal amplitudes from calorimeter cells. These amplitudes are used 
to estimate the real energy deposition of electromagnetic shower in the 
calorimeter cells by using the independent on energy calibration coefficients. 
The sum of the deposited energies in the cluster cells defines the 
energy of the incident photon or electron. This direct energy 
estimation of electromagnetic showers might be satisfactory in the energy 
range used for the calorimeter calibration but could lead to energy shifts 
at different energies which results in the calorimeter response nonlinearities 
caused by the physical processes, read-out electronics 
and shower reconstruction program. 

%
The longitudinal electromagnetic shower profile (electromagnetic cascade 
in the calorimeter radiators) \cite{PDG2010} allows to determine the shower energy
deposition in the calorimeter radiators for the case of its finite longitudinal 
thickness. The position of the energy maximum moves further into the calorimeter 
with the logarithm of the photon energy, that increases the shower energy leakage 
out of the calorimeter. Another phenomenon of the measured shower energy loss 
is related to the finite attenuation length for Cherenkov or scintillation 
light in the calorimeter cells.
The average light path from a radiation point to a photo-detector depends on
the energy of the incident photon and reveals itself also as
the nonlinear dependence with energy of the light pulse produced by shower.  
The shower energy leakage is possible in the transversal directions as well, 
for instance, due to energy loss in gaps between calorimeter cells.
 

Chosen calorimeter design could bring the nonlinearity effects as 
well. For instance, the used photo-detectors could have a nonlinear scale.
The read-out electronics (including the analog to digit converters, ADC)
could be too noisy, and the noise has to be suppressed by applying  
the relevant threshold on recorded amplitudes in the calorimeter cells.  
This threshold leads sometimes to a significant distortion of measured amplitudes 
of the incident photons at low energies. The enumeration could be continued.
But it is important to note that all these effects are unlikely possible 
to take into account with a high accuracy using Monte Carlo simulations only.
Anyway this is sufficiently difficult. 


The typical task solving by electromagnetic calorimeters 
in high energy physics experiments is the mass spectra measurement
of neutral mesons decaying into photons, for instance,
$\pi^0\to\gamma\gamma$, $\eta\to\gamma\gamma$, $\omega\to\pi^0\gamma$ and so on.
The calorimeter energy scale nonlinearity have an impact on dependence of
the measured neutral meson masses on their energies which 
leads, in turn, to systematic uncertainties in the meson spectra measurement.  
Therefore the correction of the calorimeter non-linearity response 
is relevant in the case.

At the same time the possibility of solving this problem directly 
is no means always the case, i.e. the experimental study of the calorimeter 
response to photons or electrons at different energies cannot
be carried out, for example, at collider experiments or for other reasons.
However the correction factor of energy scale of electromagnetic calorimeters 
could be found as the result of inverse problem solution, i.e. by using 
experimentally measured mass dependence of neutral mesons on the energy 
of decay photons.


In the present paper the mathematically strict algorithm of 
nonlinearity correction of the calorimeter energy scale 
based on the minimum squared deviation method is proposed. This algorithm 
has been developed and applied for the data processing from the electromagnetic 
calorimeter LGD2 of the experiment Hyperon-M at the U-70 accelerator 
of IHEP, Protvino \cite{hyperon}. 
The two photon decays of neutral pions recorded in the experiment 
have been used for the energy correction procedure. The performed 
correction allows to reduce significantly the nonlinearity of the LGD2 
energy scale and to decrease systematic uncertainties in particle mass measurement 
in several times. It opens up the possibility to obtain the
interesting physics results as well. 

It is worth to note also that the events of two photon decays of neutral mesons 
are used for a calibration purpose of the relevant electromagnetic calorimeters
in several experiments. And thus the described below procedure of 
the energy scale correction could be interesting for the data treatment in 
these experiments as well.

\section{Experiment Hyperon-М}

\begin{figure}[thb]
  \begin{center}
  \includegraphics[width=0.9\textwidth]{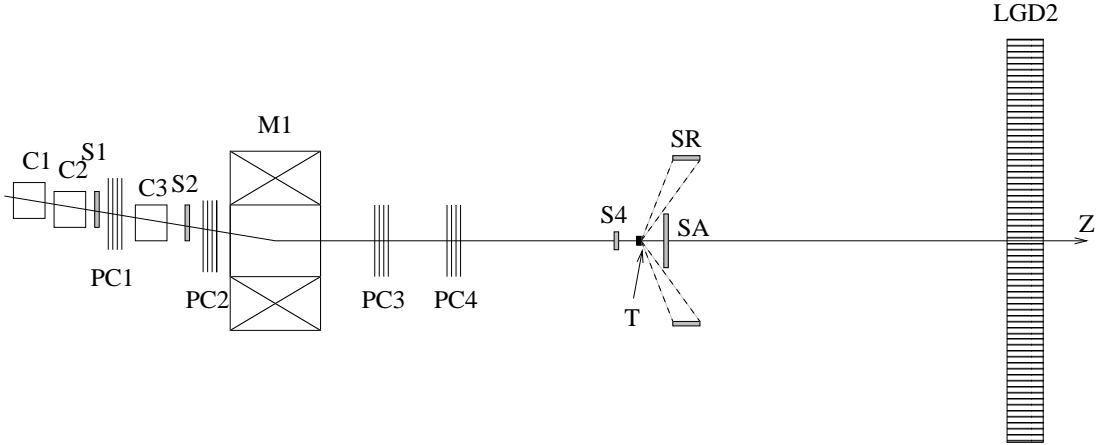}
  \end{center}
  \caption{The Hyperon-M experimental setup layout: $S_{1},~
    S_{2},~S_{4}$ --- beam scintillation counters, $C_{1-3}$ ---
    Cherenkov counters, $T$ --- nuclear target, $S_A$ --- trigger scintillation
    anti counter $S_A$, $PC_i$ --- proportional chambers, 
    LGD2~ --- Cherenkov electromagnetic calorimeter with lead glass radiators.}  
\label{fig:Hypern}
\end{figure}

It is appropriate at first to give a short description of the Hyperon-M 
setup before discussing the electromagnetic calorimeter LGD2 
energy scale in the experiment. 
The layout of experiment is presented in Fig.\ref{fig:Hypern}. 
The setup comprises the beam telescope of scintillation
counters $S_{1},~ S_{2},~S_{4}$, Cherenkov counters $C_{1-3}$, nuclear  
target $T$, scintillation anti counter $S_A$ and electromagnetic 
Cherenkov lead glass calorimeter LGD2 located at a distance of 3.7~m 
after the target. The measurements were carried out on the 7~GeV/c beam of 
positive particles with intensity of $\sim 10^{6}$~particles per burst
on different nuclear targets, including the $Be$ target.
The requirement of a beam particle signal from the beam telescope and 
the absence of a signal from anti counter $S_A$ generates the trigger 
signal:
$$
Tr = S_{1} \cdot S_{2} \cdot S_{4} \cdot \bar{S_A}.
$$
This trigger allows to select effectively the inclusive production of  
neutral mesons $M^{0}$ decaying into photons
within the LGD2 calorimeter solid angle: 
\begin{equation}
\pi^{+}(K^{+},p) \; + \;A_{z} \to M^{0} + X, \;\;\;\; M^{0} \to n \gamma.
\label{react}
\end{equation}
A typical value of the trigger selectivity reached the value of
$\sim 1$-$3 \cdot 10^{-2}$ depending on the type and thickness of the
irradiated target and the beam intensity.  More detailed description of the
Hyperon-M setup, electronics, trigger and data acquisition system can
be found elsewhere \cite{Hyp_DAQ}.


The LGD2 calibration was performed on the physics two-photon events 
collected on the $Be$ target. The sample of calibration events comprises of 
2 millions events (\ref{react}) with the reconstructed photon multiplicity 
$n=2$ and the photon pair energy $E_{\gamma\gamma} > 1.5$~GeV. 
Determination of the calibration coefficients was performed 
by means of the iterative corrections of the $\pi^0$-peak position 
with a smooth background in each calorimeter cell on the subset of 
two photon events where one of two photons hits this cell, 
details see in \cite{Blick}. We note here only that 
the effective mass of photon pair was evaluated with the formula:
\begin{equation}
m_{2\gamma} \; = \; \sqrt{2 \varepsilon_1 \varepsilon_2 (1 - \cos\theta_{12})},
\label{m2g}
\end{equation}
where $\varepsilon_i$ is the measured energy of the $i$-th photon and
$\theta_{12}$ is the opening angle of photon pair in the laboratory
frame.  The effective mass spectrum of photon pairs in reaction
(\ref{react}) after 15 iterations is illustrated by
Fig.~\ref{fig:spectr}. The obtained mass resolution for the
$\pi^0$-meson is equal to 11.4 MeV.

\section{The calorimeter energy scale correction procedure}


Let's define the nonlinear correction to the calorimeter LGD2 energy scale
$\Delta \varepsilon$ as the difference between the ``true''$ $ photon energy 
$\tilde\varepsilon$ and its measured value $ \varepsilon$:
\begin{equation}
\Delta \varepsilon = \tilde\varepsilon - \varepsilon.
\end{equation}
This correction can be expanded in a power series over some variable $x$ 
depending on the photon energy
\begin{equation}
  \Delta \varepsilon = \sum_{i=0}^{i=k}\alpha_i\cdot x^i
  \label{correc}
\end{equation}
taking into account 
that the correction $\Delta \varepsilon$ should be comparatively 
small with respect to the measured photon energy. 
To avoid the computational precision limitations at large energy values related 
to the factorisation order $k$ in expression (\ref{correc}),
it is natural to take for the $x$ variable
the logarithm of measured photon energy:
\begin{equation}
  x=x(\varepsilon)=\ln(\varepsilon/\varepsilon_0), 
\label{x}
\end{equation}
where $\varepsilon_0 = 1$~MeV. As a consequence the corrected photon energy 
$\tilde \varepsilon$ can be written as: 
\begin{equation}
\tilde \varepsilon(\varepsilon) = \varepsilon + \Delta \varepsilon = \varepsilon~ (  
                    1 + \sum_{i=0}^{i=k}{{\alpha_i}\over{\varepsilon}} ~ x^i),
\label{correc2}
\end{equation}
where it is natural to assume that the parameters $\alpha_i/\varepsilon$ 
are sufficiently small due to a small nonlinearity of the calorimeter energy scale.
The expression for the effective mass of a photon pair (\ref{m2g}) can be
rewritten then in terms of the corrected energies of photons as follows:
\begin{equation}
\widetilde m_{2\gamma} \; = \; \sqrt{2 \tilde \varepsilon_1 \tilde
  \varepsilon_2 (1 - \cos \theta_{12})} \; = \;  \sqrt{ \tilde
  \varepsilon_1 \tilde \varepsilon_2} \cdot c_{12},
\label{tm2g}
\end{equation}  
where $\tilde\varepsilon_i= \tilde\varepsilon(\varepsilon_i)$ are  
linear functions (\ref{correc2}) of small parameters $\alpha_i/\varepsilon$ 
and $c_{12}=\sqrt{1 - \cos \theta_{12}}$ is the geometrical factor which is 
actually independent on these parameters.

\begin{figure}[h*]
  \begin{center}
  \epsfxsize=9.0cm \epsfysize=6.5cm \epsfbox[100 20 330 190]{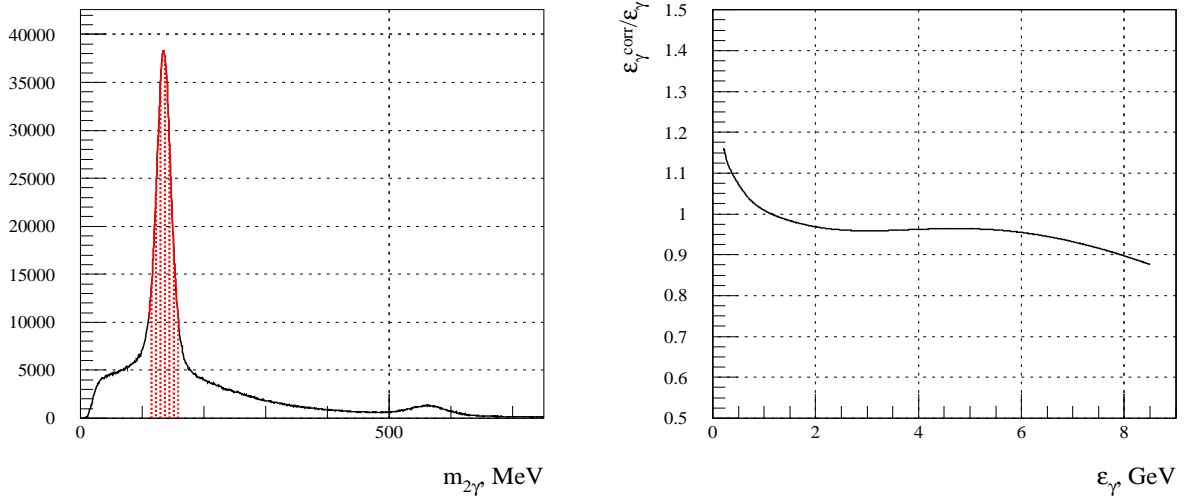}
  \end{center}
\caption{Left: the effective mass spectrum of two-photon events with 
the sum energy of photons larger than 1.5 GeV. Right: the energy scale 
correction function for LGD2 calorimeter defined on the $\pi^0$ event sample
and shown as hatched area in picture on left.}
\label{fig:spectr}
\end{figure}

The parameters $\alpha_i$ in equation (\ref{correc2}) can be determined by 
minimisation of the deviation 
of effective mass of the photon pair in representation (\ref{tm2g})
from the PDG $\pi^0$-meson mass on the sample of $\pi^0$ events used 
in the discussed procedure and  
shown for our case in Fig.\ref{fig:spectr} (left) as hatched area.
In other words, the parameters $\alpha_i$ can be determined 
by means of the functional minimisation
\begin{equation}
\chi^2= \sum_{n=1}^N {{(\widetilde m_{2\gamma} - m_{\pi^0})^2} \over
  {\sigma^2 (m_{2\gamma})}},
\label{CHI}
\end{equation}
where $N$ is the number of two-photon events in the indicated $\pi^0$-peak 
region in Fig.\ref{fig:spectr}, $\widetilde m_{2\gamma}$ is the effective 
mass of a photon pair in the representation (\ref{tm2g}), $m_{\pi^0}$
is the PDG value of the 
$\pi^0$-meson mass and $\sigma(m_{2\gamma})$ is the expected uncertainty 
of the effective pair mass as defined in expression (\ref{m2g}). 

The uncertainties on the invariant mass of photon pair include the 
photon energy uncertainty and the uncertainty of the photon pair opening angle, 
see (\ref{m2g}). The opening angle error is defined by the Hyperon-M 
setup geometry and the reconstruction program of LGD2 calorimeter.
This error is sufficiently small in our case, and we will neglect it
below. 
 
The relative uncertainty of the photon energy measurement in
electromagnetic calorimeter is defined according to the formula:
$$
\sigma_\varepsilon/\varepsilon=a/\sqrt{\varepsilon}\oplus b\oplus c/\varepsilon,
$$ 
where parameters $a$, $b$ and $c$ are defined by the calorimeter design, see 
for example \cite{PDG2010}. The last summand contribution in the energy 
resolution of LGD2 calorimeter is small and we will ignore it below. 
Thus the expected mass resolution for photon pairs can be express using the
error propagation techniques as:
\begin{equation}
\sigma^2 (m_{2\gamma}) = A(c_{12}^2 (\varepsilon_1 + \varepsilon_2) + B),
\label{msig}
\end{equation} 
where the energies of photons are measured in GeV, and $A$ and $B$ are 
the empirical parameters equal to $2.5\cdot 10^{-3}$~GeV and 
$1.4\cdot 10^{-3}$~GeV respectively for the LGD2 spectrometer.


The necessary conditions for functional (\ref{CHI}) minimisation 
$$
\partial\chi^2/\partial\alpha_i = 0
$$
with the accuracy up to the second order smallness $\alpha_i \alpha_j/\varepsilon^2$
result in the system of linear equations relatively to the parameters $\alpha_j$:
\begin{eqnarray}
\label{GF}
\sum_{j=0}^{k}\alpha_j\sum_{n=1}^N{c_{12}^2 \over 2\varepsilon_1
  \varepsilon_2 \sigma^2(m_{2\gamma})} 
(\varepsilon_1 x_2^i +\varepsilon_2 x_1^{i})(\varepsilon_1 x_2^j+\varepsilon_2 x_1^j) = \\
 =\sum_{n=1}^N({m_\pi^0\over\sqrt{\varepsilon_1
     \varepsilon_2}}-c_{12}){c_{12} \over
   \sigma^2(m_{2\gamma})}(\varepsilon_1x_2^i+\varepsilon_2 x_1^i),\nonumber\\
\nonumber
\end{eqnarray}
where $x_l = x(\varepsilon_l), ~ l=1,2$, see equation (\ref{correc2}).
The iteration procedure based on equations (\ref{correc2}) and (\ref{GF}),
allows one to find out the functional minimum (\ref{CHI}) with a reasonably 
good accuracy after $2-3$ iterations.

 
The first 9 terms of the series (\ref{correc2}), i.e up to the order of $k=8$, 
were taken into account for the Hyperon-M data treatment. The next values 
for the correction coefficients have been obtained after the first iteration:
~$\alpha_{0-8}/${\it GeV}~ = ~0.00399,~~ -0.0505,~~ -0.0392,~~ -0.0209,~~ 
-0.00537,~~  0.0165,~~  0.0104, ~~-0.00313,~ -0.00235.
The coefficients of the second iteration were found to be about three
times less compared with the first iteration values. The energy correction function 
for the LGD2 calorimeter  
\begin{equation}
%
\varepsilon^{corr}/\varepsilon = \tilde\varepsilon(\varepsilon)/\varepsilon =
1 + \sum_{i=0}^{i=k}{{\alpha_i}\over{\varepsilon}} ~ x^i
\label{fcor}
\end{equation} 
is presented in Fig.\ref{fig:spectr} in the right plot.  As one can
see from the figure the relative correction doesn't exceed 10\% level
virtually in the whole energy range of photons and this is in a good
agreement with our initial assumption concerning the smallness of an
expected energy scale nonlinearity of the LGD2 calorimeter. This is an
important statement because it is used in the ground of the method.
For the sake of completeness it would be useful also to present the
values of $\chi^2$ (\ref{CHI}) before and after the correction: in our
case the value of $\chi^2$ per degree of freedom before the
correction and after it are equal to $1.073$ and $1.044$ respectively
for approximately $10^6$ degrees of freedom.

\section{Results and discussion}

\begin{figure}[htb]
  \begin{center}
   \epsfxsize=8.0cm \epsfysize=8.0cm \epsfbox[60 30 190 190]{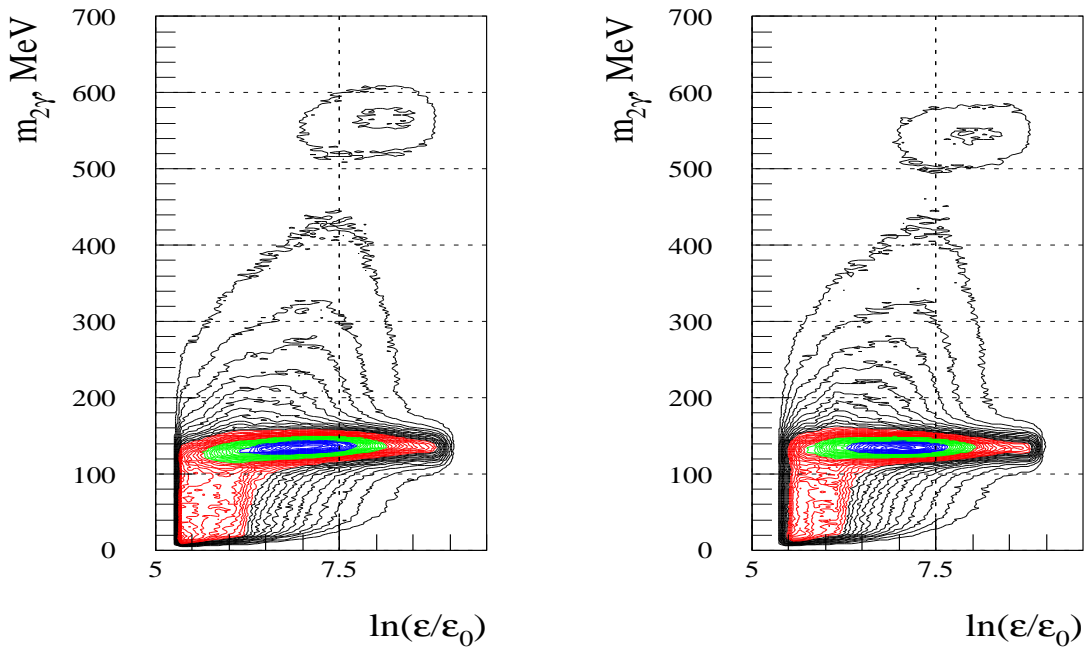}
\end{center}
\caption{Distributions of the two-photon effective mass versus the 
logarithm of photon energy of each photon in pair (two points per event) 
for the reconstructed two-photon events on Be-target. The concentration 
of events at lower area corresponds to the detection of $\pi^0$-mesons, 
upper one -- to the $\eta$-mesons. The event distributions before the energy 
scale correction are shown on left and after the correction -- on right.}
\label{fig:depth}
\end{figure}

Performance of the above discussed procedure is illustrated in Fig.\ref{fig:depth},
where the scatter plots of the effective two photon mass versus the logarithm of the
energy of each photon in a pair (two points per event) for the reconstructed 
two-photon events (\ref{react}) is shown for $Be$-target before the energy scale 
correction on the left panel and after it on the right panel of the figure. 
A clear correlation of the two-photon mass and the photon energies for events 
in the $\pi^0$-meson region is seen on the left picture and it is completely 
absent on the right one. The numerical values of the
correlation coefficients for events without the energy scale correction and 
with it are equal to 0.13 and 0.05 respectively. 


\begin{figure}[hbt]
  \begin{center}
  \epsfxsize=7.5cm \epsfysize=5.6cm \epsfbox[110 30 330 190]{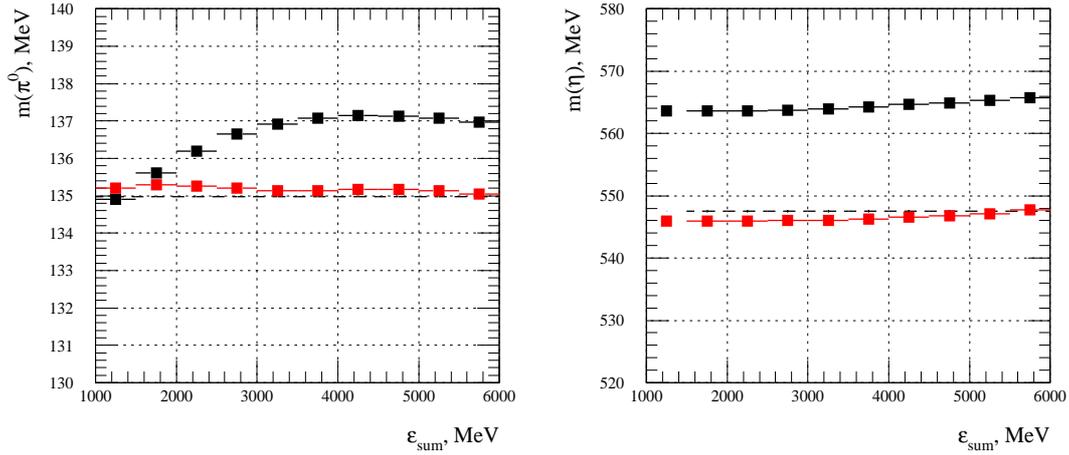}
  \end{center}
  \caption{The mass dependence of $\pi^0$-meson (on left) and $\eta$-meson
    (on right) versus the minimal photon pair energy
    $\varepsilon_{\rm sum}=\varepsilon_1+\varepsilon_2$.  The dependence
    before the LGD2 energy scale correction is shown by black colour and
    one after correction is shown by red colour. The dashed line shows
    the PDG values of these mesons.}
\label{fig:pi-et}
\end{figure}
Another illustration of the nonlinearity correction method is
represented by Fig.~\ref{fig:pi-et}. The dependence of the measured
mass of the $\pi^0$- and $\eta$-mesons on the minimal photon pair
energy ($\varepsilon_{\rm sum}=\varepsilon_1+\varepsilon_2$) is shown
before applying the nonlinearity correction and after it. These plots demonstrate as
well that the systematic deviation
of the neutral pion mass from the PDG value in dependence on the
photon pair energy decreases from 1.17\% to 0.19\%, i.e. in 6 times,
and the same deviation for $\eta$-meson decreases from 2.98\% to
0.23\%, i.e. in 13 times, and this is demonstration of the high
performance of the proposed method as a whole.

\section*{Conclusion}

This paper describes the procedure of the energy scale correction for electromagnetic 
calorimeters. The procedure is based on the minimization of mass resolution 
for two-photon decays of neutral pion detected in the calorimeter.
The linear parametrisation of the correction function as the power series in 
logarithm of the photon energy allows to provide a simple and effective
energy scale correction in a very wide energy range.
Possibility to use the physics statistics of the experiment 
for the energy correction procedure results in the high accuracy and sensitivity
of the method. For instance, in the Hyperon-M experiment the reached mass scale 
nonlinearity for two-photon events is equal to 0.2\%, that hardly can be 
obtained in calculations of similar corrections by Monte-Carlo methods due to 
restrictions peculiar to the transport code. Anyway the significant Monte-Carlo 
difficulties appear in calculations at the accuracy level of $10^{-3}$. 

Eventually, it is significant that the described procedure could be applied
practically for any hodoscopic electromagnetic calorimeter if the physics 
statistics of experiment possesses the needed amount of two-photon or, let's say,
three-photon decays of known mesons because this procedure could be easy 
generalised for multi-photon decays as well.

\vskip 0.2cm
The authors appreciate N.A.Kuzmin for discussions and helpful comments.

%

\hskip 11.0cm {\it Received December 22, 2010.}

\end{document}